# An investigation of the in-plane dc fluctuation conductivity of optimally doped and overdoped cuprates: implication and origin of the pseudogap

S H Naqib[*] and R S Islam

*Department of Physics, University of Rajshahi, Rajshahi 6205, Bangladesh*
*Corresponding author. E-mail: salehnaqib@yahoo.com

**Abstract**

In conventional superconductors the magnitude of the pairing fluctuation is primarily determined by $T_c$ and the superconducting coherence length, $\xi$. In low-dimensional systems with strong structural and electronic anisotropies, the interlayer separation, $s$, plays a significant role. In cuprates the pseudogap (PG) correlation induces a downturn in the temperature dependent resistivity. As $T_c$ is approached from above, this downturn in the resistivity is supposed to either i) add to or ii) join smoothly to that due to paraconductivity caused by short-lived Cooper pairs. It is important to differentiate between these two possibilities since they are closely linked to the origin of the PG. It would be reasonable to assume that if the first scenario is correct then the PG has a non-superconducting origin, while the second scenario, if found to hold, would relate precursor pairing to the PG correlations. In this paper we have studied the in-plane fluctuation conductivity of two $c$-axis oriented thin films of $Y_{0.95}Ca_{0.05}Ba_2Cu_3O_{7-\delta}$ with similar hole contents ($p$), $p = 0.165$ (optimally doped) and $p = 0.184$ (slightly overdoped). The hole contents are fixed at these values so that the PG affects the slope of the resistivity data only at temperatures close to $T_c$. Analysis of paraconductivity, $\Delta\sigma_{ab}(T)$, of these compounds within the mean-field Gaussian Ginzburg-Landau (MFGGL) framework reveals different qualitative and quantitative features for the optimally doped (OPD) and the slightly overdoped (SOD) compounds. The excess conductivity due to Cooper pair fluctuations of the SOD sample can be described reasonably well by the MFGGL formalism. The excess conductivity of the OPD compound, on the other hand, cannot be accounted for by the MFGGL formalism with reasonable set of parameters. There is a significant added contribution to $\Delta\sigma_{ab}(T)$ for the OPD sample which appears to come from the presence of a PG in the quasiparticle (QP) spectral density. These findings point towards a non-pairing origin of the PG.

## 1. Introduction

Ever since the discovery of high-$T_c$ cuprates, there has been intense debate on the question of how the parent Mott antiferromagnetic (AFM) insulator transforms itself in to a high-$T_c$ superconductor upon carrier doping, but the answer remains elusive till date. The situation is complicated because these remarkable materials exhibit various normal state correlations (which may well exist in the superconducting state) as the number of added holes, $p$, in the $CuO_2$ plane is varied. Besides superconductivity itself, the most extensively studied phenomenon is undoubtedly the pseudogap (PG) in the quasiparticle (QP) energy spectrum [1 – 5]. Depending on the experimental probe and the method of data analysis, PG either goes below the superconducting (SC) dome and falls to zero abruptly at a critical doping in slightly overdoped (OD) side, or it approaches and tracks the superconducting transition temperature in the deeply OD region [1 – 3, 6 – 8]. These two alternatives have profound physical significances. The former suggests that PG is unrelated to SC pairing fluctuations, while the later links the PG with the precursor pairing. To build a successful microscopic theory for superconductivity in cuprates, it is essential to scrutinize the two scenarios of PG. The nature of the dominant normal state correlations must be pinned down accurately since SC itself emerges from it as temperature is lowered.

In some earlier studies on in-plane paraconductivity of hole doped cuprates [9 – 12] the link between SC fluctuations and the PG has been investigated. Conflicting results can be found in the literature, for example, Leridon *et al.* [9] interpreted their analysis in terms of incoherent pairing fluctuations in the PG regime. In this particular study a heuristic expression was used for the paraconductivity at high reduced temperature, $\varepsilon$ ($\equiv \ln(T/T_c)$). Vidal *et al.* [10, 11] on the other hand, used the conventional Gaussian Ginzburg-Landau (GGL) formalism with proper total energy cut-off to study the fluctuation conductivity in the high $\varepsilon$ region. These studies found that the onset of the PG does not affect the magnitude of the in-plane paraconductivity, $\Delta\sigma_{ab}(\varepsilon)$ in any significant way. Such conclusion is also supported by the study by Albenque *et al.* [12]. It is interesting to note that the apparent insensitivity of $\Delta\sigma_{ab}(\varepsilon)$ on the onset and presence of the PG in the QP energy spectrum cannot distinguish between the proposed scenarios for the origin of the PG in any conclusive fashion. A lot depend on the way the background conductivity has been determined. For example if the enhanced conductivity (the downturn of resistivity from its high-$T$ linear behaviour much above $T_c$) due to the PG is already included in the background, then the extracted SC fluctuation contribution to the conductivity would be just due to precursor pairing and it may appear that $\Delta\sigma_{ab}(\varepsilon)$ is indeed independent of the presence of the PG. To avoid this from happening, one has to be careful about the determination of the background conductivity from the high-$T$ fits of the resistivity data.

In this paper we wish to distinguish between two scenarios – i) the enhanced conductivity due to PG merges smoothly to the mean-field (Gaussian) paraconductivity near $T_c$ and has the same character ii) PG induces an additional and independent contribution to the Gaussian paraconductivity of different nature. We assume that, if the first scenario is found to be correct then PG will be linked with precursor pairing. While for the second scenario, PG originates from correlations of non-SC nature.

The number of added holes in the $CuO_2$ plane determines almost all the normal and SC state properties of high-$T_c$ cuprates. The profile of the temperature dependent resistivity is no exception. The temperature range over which the in-plane resistivity, $\rho_{ab}(T)$, is linear grows progressively with hole content and the characteristic downturn due to PG shifts to lower and lower temperatures [1, 13, 14]. In the deeply OD side $\rho_{ab}(T)$ exhibits a superlinear temperature dependence except at temperatures quite close to $T_c$ [1, 13]. In this study we have analyzed the excess conductivity of two high-quality $Y_{0.95}Ca_{0.05}Ba_2Cu_3O_{7-\delta}$ (Y(Ca)123) thin films with $p = 0.165$ (optimally doped(OPD)) and 0.184 (slightly overdoped (SOD)). From earlier analysis of the $\rho_{ab}(T)$ data, the characteristic PG temperatures ($T^*$) for these hole contents were found to be $135 \pm 10$ K and $45 \pm 10$ K, respectively [14]. The lower value ~ 45 K, below the $T_c(p)$ dome of pure Y(Ca)123 has been obtained by suppressing superconductivity by Zn [14]. If one prefers to take the onset of strong SC fluctuations near $T_c$, where $\rho_{ab}(T)$ shows a sharp and accelerating downturn as an indication for the PG, then $T^*$ becomes $113 \pm 10$ K [13, 14] for the SOD compound under consideration. In either case, for both the hole contents under study, $\rho_{ab}(T)$ is completely linear above 150 K. Therefore, a linear fit to $\rho_{ab}(T)$ above 150 K will not be affected by the PG in the QP energy spectral density, which for the selected compositions reveals itself at lower temperatures.

By definition, the background conductivity should be free from paraconductive contributions. Vidal *et al*., [10, 11, 15] have developed and successfully implemented mean-field Gaussian Ginzburg-Landau (MFGGL) framework incorporating a total energy cut-off to analyze the fluctuation conductivity of different families of cuprates. An interesting feature of this formalism is the concept of a cut-off in the reduced temperature above which short-lived Cooper pairs cease to exist as both amplitude and phase fluctuations die down. This follows from the consideration of the uncertainty principle and the size of the fluctuating pairs as thermal energy is increased. It was shown in ref. [15] that paraconductive contributions vanish abruptly at reduced temperatures above $\varepsilon \sim 0.50$. Superconducting transition temperatures of the experimental samples used in this paper lie within the range $90 \pm 1$ K.

Therefore, a background fit to $\rho_{ab}(T)$ data for $T > 150$ K is expected to be free from both PG and SC fluctuations.

In subsequent sections we have analyzed the paraconductivity of $Y_{0.95}Ca_{0.05}Ba_2Cu_3O_{7-\delta}$ films using the MFGGL formalism. From this analysis it appears that PG adds to an additional contribution to $\Delta\sigma_{ab}(\varepsilon)$ for the OPD compound which cannot be taken into account by the MFGGL treatment even with grossly unreasonable set of fitting parameters. $\Delta\sigma_{ab}(\varepsilon)$, on the other hand, for the SOD sample is significantly smaller and can be better modelled by the MFGGL formalism. This finding supports the scenario where PG and SC correlations have different physical origin.

## 2. Experimental samples and results

Crystalline *c*-axis oriented thin films of $Y(Ca)Ba_2Cu_3O_{7-\delta}$ were synthesized from high-density single-phase sintered targets using the method of pulsed laser deposition (PLD). Ca substitution enabled us to access the OD side relatively easily. Thin films were grown on (*001*) SrTiO$_3$ single crystal substrates at a deposition temperature of 800 °C under an oxygen partial pressure of 1.0 mbar. Samples were characterized by using X-ray diffraction (XRD), atomic force microscopy (AFM), *ab*-plane room-temperature thermopower, $S_{ab}[290\ K]$, and $\rho_{ab}(T)$ measurements. XRD was used to determine the lattice parameters, phase-purity, and degree of *c*-axis orientation (extracted from the rocking-curve analysis). XRD profile of the OPD $Y_{0.95}Ca_{0.05}Ba_2Cu_3O_{7-\delta}$ is shown in Fig. 1. AFM was employed to study the grain size and the thickness of the films. All the films used in the present study were phase-pure (within the resolution of the X-ray diffractometer) and had high-degree of *c*-axis orientation (typical value of the full-width at half-maximum of (*007*) peak was ~ 0.28°). Thickness of the films lied within the range (2800 ± 300) Å. The hole concentrations were changed by changing the oxygen deficiency in the CuO$_{1-\delta}$ chains by annealing the films under different temperatures in different oxygen environments. $S_{ab}[290\ K]$ was used to calculate the planar hole concentration following the method proposed by Obertelli *et al.* [16]. The level of oxygen deficiency was determined from an earlier work where the relation between $\delta$ and $p$ was established [17]. Also, as an independent check, systematic variations in the *c*-axis lattice parameters were noted as $\delta$ changes as a result of oxygen annealings [17]. The hole contents quoted in this paper are accurate within ± 0.005. $\rho_{ab}(T)$ was measured on pattern films using low resistance contacts. Resistivity measurements gave additional information regarding the impurity content and about the quality of the grain-boundaries of the films. All the films used in this study have low values of $\rho_{ab}(300\ K)$ and the extrapolated zero temperature resistivity, $\rho_{ab}(0\ K)$. The low value of the residual resistivity indicates that the samples can be treated in the clean limit while considering the various length scales, the SC coherence length, $\xi$, in particular. Details of the PLD method and film characterizations can be found in refs. [17, 18]. The in-plane resistivities of the samples used in this study is shown in Fig. 2. It should be noted that 5% Ca substitution does not change any structural or electronic property of pristine Y123 in any significant way, other than facilitating one to access the OD side relatively easily. It is quite difficult to achieve full oxygenation in Y123 (which is required to raise $p$ values above 0.18 in pristine Y123) [19 - 21].

## 3. Analysis of fluctuation conductivity

The experimental in-plane fluctuation conductivity, $\Delta\sigma_{ab}(T)$, is taken as, $\Delta\sigma_{ab}(T) = [1/\rho_{ab}(T) - 1/\rho_{bg}(T)]$, where $\rho_{bg}(T)$ is the background resistivity, i.e., the resistivity in the absence of paraconductive contributions. Fig. 3 shows the fitted $\rho_{bg}(T)$ together with the experimental resistivity. The experimentally determined fluctuation conductivity is shown versus reduced temperature in Fig. 4.

For theoretical analysis, one needs to locate the mean-field superconducting transition temperature. We have taken $T_c$ at the peak temperature of the $d\rho_{ab}/dT$ data. The full width at half maximum gives a measure of the transition width. $T_c$ of the SOD compound is 89.7 K and that for the OPD one is 90.8 K. The transition widths are 1.0 K and 0.8 K for the SOD and the OPD samples, respectively. Paraconductivity is conventionally studied within the mean-field Aslamazov-Larkin theory. The $d$-wave order parameter and the energy scale associated with high $T_c$ itself leads to strong pair-breaking and therefore, the Maki-Thompson contribution is usually omitted [22]. Within the MFGGL scheme, the fluctuation conductivity arises from the fluctuating order parameter. This excess conductivity is controlled by temperature dependent SC coherence length, $\xi(T)$. Y123 (and Y(Ca)123) is a system with moderate structural and electronic anisotropy [23]. Thus over a significant temperature range above $T_c$, $\Delta\sigma_{ab}(\varepsilon)$ can be modelled by using the Lawrence-Doniach (LD) theory for anisotropic 3D superconductors [24]. In the LD scheme the paraconductivity is given by

$$\Delta\sigma(\varepsilon)^{LD} = \frac{e^2}{16\hbar s\varepsilon} \frac{1}{\sqrt{1+2\alpha}}$$

(1)

where the interlayer coupling parameter $\alpha$ is expressed as, $\alpha = 2[\xi_c(\varepsilon)/s]^2$. The temperature dependent $c$-axis coherence length, $\xi_c(\varepsilon) = \xi_c(0)/\sqrt{\varepsilon}$ [12]. $s$ is the interlayer spacing. It has been argued that for bilayered cuprates with moderate structural and electronic anisotropy, the effective interlayer periodicity should be taken as half of the *c*-axis lattice parameter [10]. Therefore we have fixed $s = 11.7/2 = 5.85$ Å for analysis of the fluctuation conductivity data.

Figs. 5 illustrate the fits of the experimental paraconductivity within the LD formalism. A reasonably good fit is obtained for the SOD compound (Fig. 5a). Eqn. 1 contains a single free parameter, $\xi_c(0)$. The extracted value of $\xi_c(0)$ from the best fit to Eqn. 1 was found to be 0.6 Å for the SOD thin film. Whereas $\xi_c(0) \sim 0.1$ Å for the OPD thin film. The quality of fit to the paraconductivity data is far inferior for this optimally doped sample (Fig. 5b). For the hole contents of the films under study, the accepted value of $\xi_c(0)$ for Y123 lies within the range $1.0 \pm 0.1$ Å [9 – 12]. The extracted value of $\xi_c(0)$ is somewhat low for the SOD compound, whereas it is an order of magnitude lower for the OPD compound. Even with this unrealistically low value of $\xi_c(0)$, the quality of fit is poor for the $p = 0.165$ sample.

LD formalism without any cut-off condition in energy or momentum, overestimates the paraconductivity in the high reduced temperatures. Vidal *et al.*, has imposed a total-energy cut-off (EC) condition, given by $[k^2 + \xi^2(\varepsilon)] < b\xi(0)^{-2}$ (in units of $\hbar^2/2m_{cp}$, where $m_{cp}$ is the effective mass of the preformed Cooper pair) [10, 15]. This forces the paraconductivity to vanish at high reduced temperatures. From theoretical ground, the value of the constant $b$ should be close to unity. This condition eliminates the high-energy fluctuating modes present at high temperatures, with a spatial extent lower than $\xi(0)$, from contributing to the fluctuation conductivity. Within this framework the paraconductivity for anisotropic 3D superconductors can be expressed as follows

$$\Delta\sigma_{ab}(\varepsilon)^{EC} = \frac{e^2}{16\hbar s}[\frac{1}{\varepsilon}(1+\frac{B_{LD}}{\varepsilon})^{-\frac{1}{2}} - \frac{2}{b} + \frac{(2\varepsilon + B_{LD})}{2b^2}]$$

(2)

$B_{LD}$ is the Lawrence-Doniach parameter that controls the fluctuation dimensionality, given by $[2\xi_c(0)/s]^2$. Fig. 6 exhibits the fit for experimental paraconductivity data where Eqn. 2 has been used, for the SOD compound. The quality of fit has improved slightly compared to that shown in Fig. 5b. More importantly, the best fit is obtained with $\xi_c(0) \sim 0.9$ Å and $b \sim 0.8$, in excellent agreement with experimentally and theoretically agreed values for the respective parameters [9 – 12, 15]. No improvement in the quality of the fit is found for the experimental paraconductivity of the OPD compound even when Eqn. 2 is employed. The best fit for this compound is of similar quality as the one shown in Fig. 5a. The values of the fitting parameters $\xi_c(0)$ and $b$, extracted from the best fit are totally unrealistic, $\sim 11$ Å and $\sim 23$, respectively.

## 4. Discussion and conclusions

From the analysis of the paraconductivity data it is evident that $\Delta\sigma_{ab}(\varepsilon)$ for the OPD compound cannot be modelled within the MFGGL scheme irrespective of theoretical details (with or without the cut-off condition). The best fits of $\Delta\sigma_{ab}(\varepsilon)$ for the OPD film yield unphysical values for the $c$-axis coherence length. $\xi_c(0)$ for the OPD and the SOD samples should be almost identical. This follows from the fact that for high-$T_c$ cuprates with $p > 0.16$ the mean-field SC transition temperature becomes almost identical to the experimental $T_c$ [25] and the canonical relation between $T_c$ and coherence length holds rigorously.

In any analysis of the paraconductivity data, there are two sources of error. The first one is related to the identification of the mean-field $T_c$. Considering the transition widths, the uncertainty in the quoted values of $T_c$ should be less than a Kelvin for the samples under study. Thus the conclusions drawn from the analysis of $\Delta\sigma_{ab}(\varepsilon)$ should not be affected by this factor. The second one is related to the identification of the background conductivity and has the potential to introduce substantial error in estimating the experimental paraconductivity. By selecting sample compositions for which $\rho_{ab}(T)$ is linear over a significant temperature range, we have been able to minimize this uncertainty in background conductivity to a large extent. Moreover, we have selected samples with hole contents such that the resistivity is linear above 150 K. This is important in two counts - i) the onset of the PG sets in below 150 K and ii) pairing fluctuations are expected to be absent above 150 K for both compounds [13 – 15]. This ensures that the extracted experimental paraconductivity includes the effect of the PG (if any) completely on the conductivity of the films under study, since background conductivity is obtained by fitting the experimental $\rho_{ab}(T)$ above 150 K over an extended temperature region.

It is important to notice that the disagreement between experimental and theoretical paraconductivity is pronounced for the OPD sample and tends to set in at $T \sim 130$ K ($\varepsilon \sim 0.35$) (Fig. 5a). This temperature is quite close to $T^*$ ($\sim 135$ K) and indeed a disagreement between MFGGL paraconductivity and experimental paraconductivity is expected when it is enforced by a significant contribution due to a PG not related to precursor pairing correlations. A small disagreement between theory and experiment sets in at $T \sim 98$ K ($\varepsilon \sim 0.09$) for the SOD sample (Fig. 6). This might be due to presence of a small PG. In any case, as the PG decreases the agreement between theory and experiment improves markedly.

It is also worth noticing that introducing a cut-off criterion in the total energy improves the agreement between theoretical and experimental fluctuation conductivities for the SOD compound. At the same time the extracted parameters turn out to be in excellent agreement with the experimentally and theoretically established values. The situation for the OPD sample does not improve. This again can be explained by invoking to the fact that the cut-off condition works for Gaussian pairing fluctuations modes, whereas the added paraconductivity-like contribution comes from a non-SC correlations in the OPD sample.

It is customary to plot the paraconductivity versus reduced temperature data using log scale. This enhances the behavior in the high reduced temperature region where paraconductivity is low. In this study we have used a linear scale throughout, since our interest was to scrutinize the overall agreement between theoretical and experimental paraconductivities over an extended temperature range. This study also indicates that both UD and deeply OD compounds are somewhat inappropriate for theoretical analysis of the extracted paraconductivity because of unavoidable uncertainty associated with determination of the background conductivity. UD compounds with large PG, develop a downturn in the resistivity at temperatures much above $\varepsilon > 0.5$. Resistivity for deeply OD compounds, on the other hand, show superlinear behaviour. In either case polynomial fits for background resistivity becomes necessary. Without any clear physical meaning of the various terms in such polynomial fits, the extrapolated background resistivity used to extract the fluctuation conductivity become dubious and can induce significant error in this quantity (paraconductivity). Linear resistivity at *high* temperatures is a generic feature for all hole doped cuprates and may be related to quantum critical fluctuations [1, 26, 27]. Linear fit over an extended temperature range becomes possible only for samples with $p$ values lying within the range $0.16 < p < 0.20$.

The apparent agreement between MFGGL fluctuation conductivity and the experimental paraconductivity below $\varepsilon \sim 0.02$ for samples with varied compositions [10 – 12] can be accounted for by realizing that, at these temperatures the contribution from pairing fluctuations are significantly higher than any contribution arising from the presence of a coexisting PG.

To summarize, we have found that the PG adds to an additional contribution to the paraconductivity over a wide temperature range. The nature of this extra conductivity is fundamentally different from that due to fluctuating Cooper pairs. As the hole content increases and the PG decreases, the paraconductivity becomes increasingly purely MFGGL type. All these reinforce the idea that PG originates from electronic correlations unrelated to that giving rise to phase incoherent precursor pairing.

**Acknowledgements**

The authors thank the Cavendish Laboratory, University of Cambridge, UK, Commonwealth Commission, UK, and Trinity College, University of Cambridge, UK, for support with experimental facilities and financial assistance.

**References**

[1] Tallon J L and Loram J W 2001 *Physica* C **349** 53

[2] Lee P A 2008 *Rep. Prog. Phys.* **71** 012501

[3] Damascelli A, Hussain Z and Shen Z –X 2003 *Rev. Mod. Phys.* **75**, 473

[4] Makoto Hashimoto *et al.* 2014 *Nature Materials* DOI: 10.1038/NMAT4116

[5] Fujita K *et al.* 2014 *Science* **344** 612

[6] Vivek Mishra, Chatterjee U, Campuzano J C and Norman M R 2014 *Nature Physics* **10** 357

[7] Emery V J and Kivelson S A 1995 *Nature* **374** 434

[8] Lee P A, Nagaosa N and Wen X –G 2006 *Rev. Mod. Phys.* **78** 17

[9] Leridon B, Defossez A, Dumont J, Lesueur J and Contour J P 2001 *Phys. Rev. Lett.* **87** 197007

[10] Carlos Carballeira, Severiano R Curras, Jose Vina, Jose A Veira, Manuel V Ramallo and Felix Vidal 2001 *Phys. Rev.* B **63** 144515

[11] Severiano R Curras, Gonzalo Ferro, Teresa Gonzalez M, Manuel V Ramallo, Mauricio Ruibal, Jose Antonio Veira, Patrick Wagner and Felix Vidal 2003 *Phy. Rev.* B **68** 094501

[12] Rullier-Albenque F, Alloul H and Rikken G 2011 *Phys. Rev.* B **84** 014522

[13] Naqib S H, Cooper J R, Tallon J L, Islam R S and Chakalov R A 2005 *Phys. Rev.* B **71** 054502

[14] Naqib S H, Cooper J R, Tallon J L and Panagopoulos C 2003 *Physica* C **387** 365

[15] Vidal F, Carballeira C, Curras S R, Mosqueira J, Ramallo M V, Veira J A and Vina J 2002 *Europhys. Lett.* **59** 754

[16] Obertelli S D, Cooper J R and Tallon J L 1992 *Phys. Rev.* B **46**, 14928

[17] Naqib S H and Semwal A 2005 *Physica* C **425** 14

[18] Naqib S H 2003 *Ph.D. thesis* University of Cambridge, UK (unpublished)

[19] Naqib S H, Chakalov R A and Cooper J R 2004 *Physica* C **407** 73

[20] Naqib S H, Cooper J R and Loram J W 2009 *Phys. Rev.* B **79** 104519

[21] Tallon J L, Bernhard C, Shaked H, Hitterman R L and Jorgensen J D 1995 *Phys. Rev.* B **51** 12911

[22] Kip S K 1990 *Phys. Rev.* B **41** 2612

[23] Lundqvist B, Rydh A, Eltsev Yu, Rapp O and Andersson M 1998 *Phys. Rev.* B **57** 14064 and Nagasao K, Masui T and Tajima S 2008 *Physica* C **468** 1188

[24] Lawrence W E and Doniach S 1971 *Proceedings of 12$^{th}$ International Conference on Low Temperature Physics* Kyoto 1970 (Edited by Kanda E) 361

[25] Tallon J L, Storey J G and Loram J W 2011 *Phys. Rev.* B **83** 092502

[26] Tallon J L, Loram J W, Williams G V M, Cooper J R, Fisher I R, Johnson J D, Staines M P and Bernhard C 1999 *Phys. Stat. Sol.* (b) **215** 531

[27] Naqib S H, Borhan Uddin M and Cole J R 2011 *Physica* C **471** 1598

**Figure captions**

Figure 1: X-ray diffraction spectra for the optimally doped $Y_{0.95}Ca_{0.05}Ba_2Cu_3O_{7-\delta}$. The Miller indices (*hkl*) and the angles (*2θ*) are indicated.

Figure 2 (color online): Temperature dependent in-plane resistivities of the $Y_{0.95}Ca_{0.05}Ba_2Cu_3O_{7-\delta}$ thin films.

Figure 3 (color online): Linear background fits (full straight lines) to the experimental in-plane resistivities of the $Y_{0.95}Ca_{0.05}Ba_2Cu_3O_{7-\delta}$ thin films.

Figure 4 (color online): Experimental paraconductivity of the $Y_{0.95}Ca_{0.05}Ba_2Cu_3O_{7-\delta}$ thin films versus reduced temperature.

Figure 5 (color online): Experimental paraconductivity (circles) and fitted paraconductivity, using the LD scheme, (full line) versus the reduced temperature for (a) the optimally doped and (b) the slightly overdoped compounds.

Figure 6 (color online): Experimental paraconductivity (circles) and fitted paraconductivity, using the total energy cut-off condition, (full line) versus the reduced temperature for the slightly overdoped compound.

Figure 1

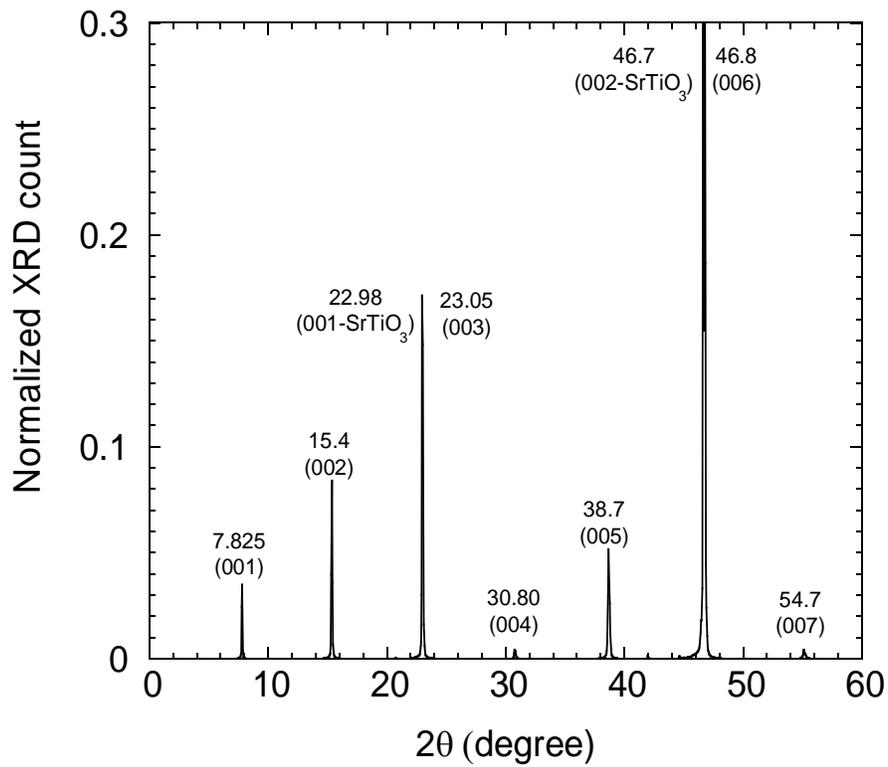

Figure 2

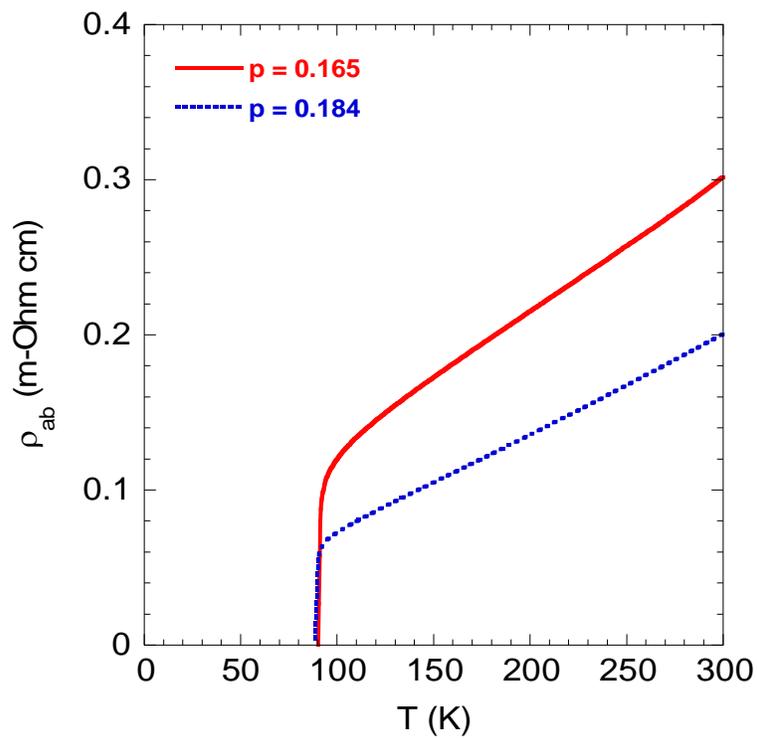

Figure 3

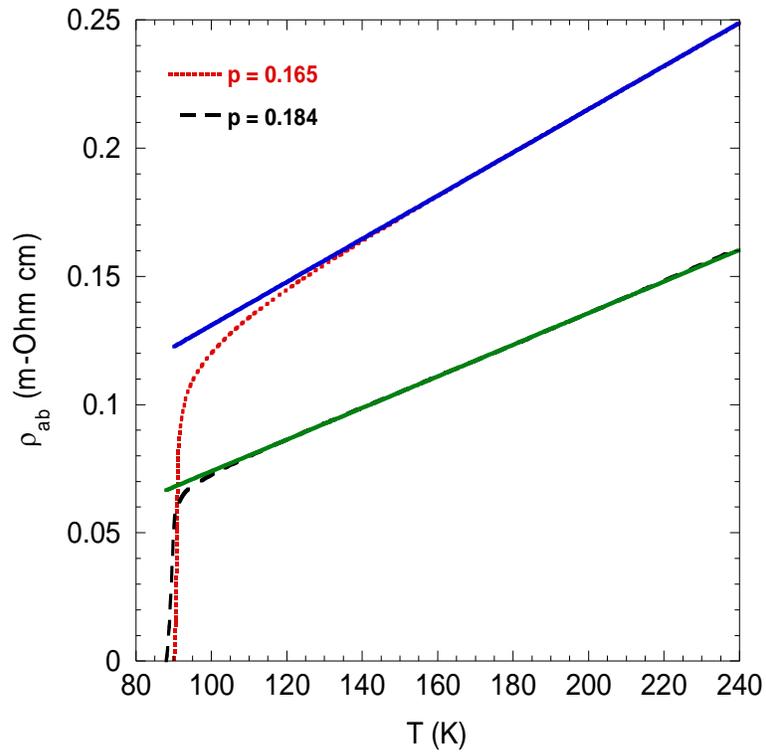

Figure 4

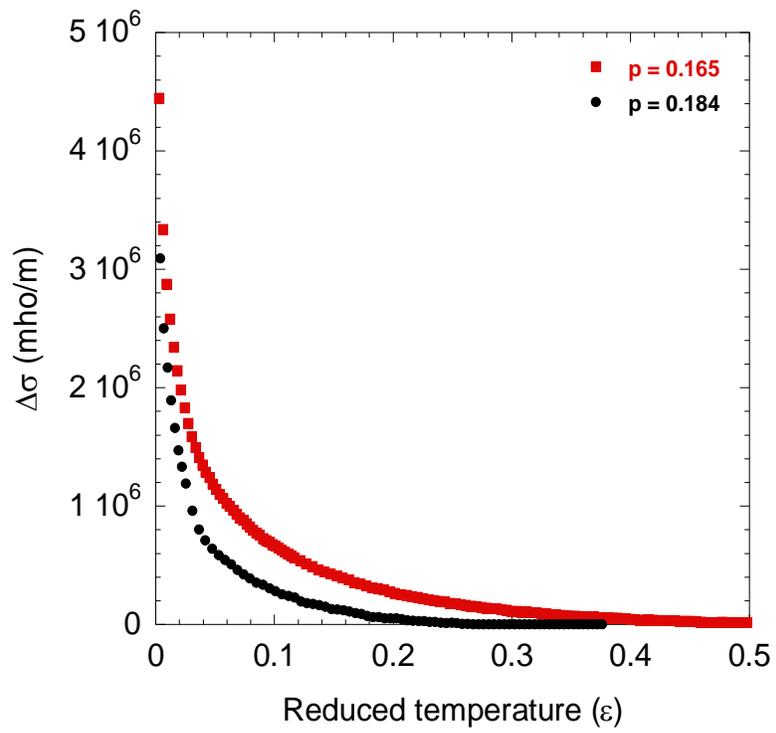

Figure 5

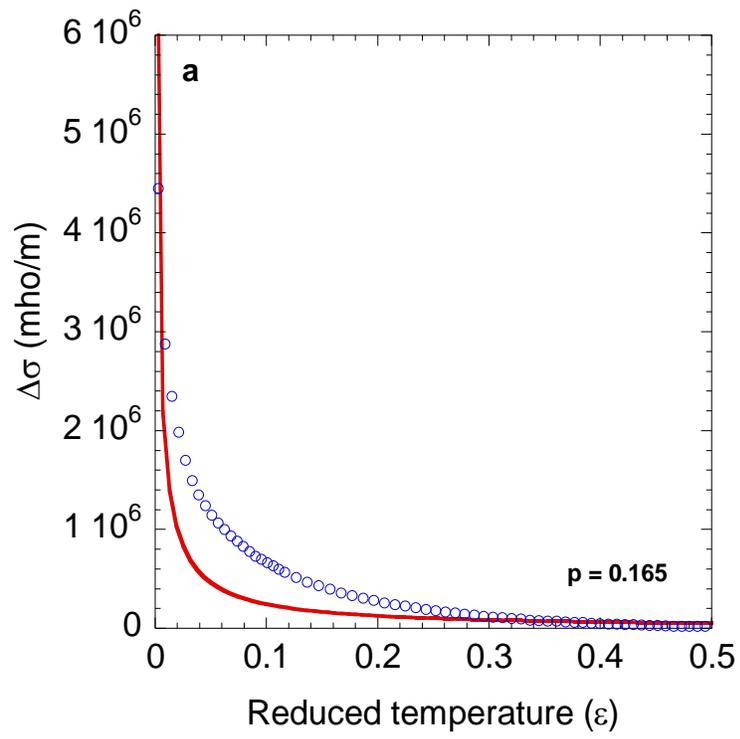

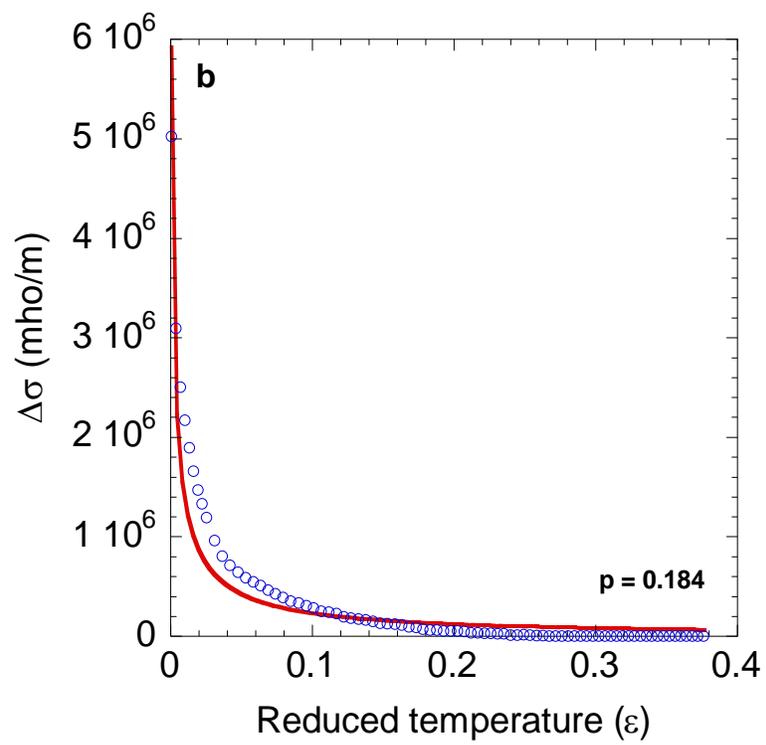

Figure 6

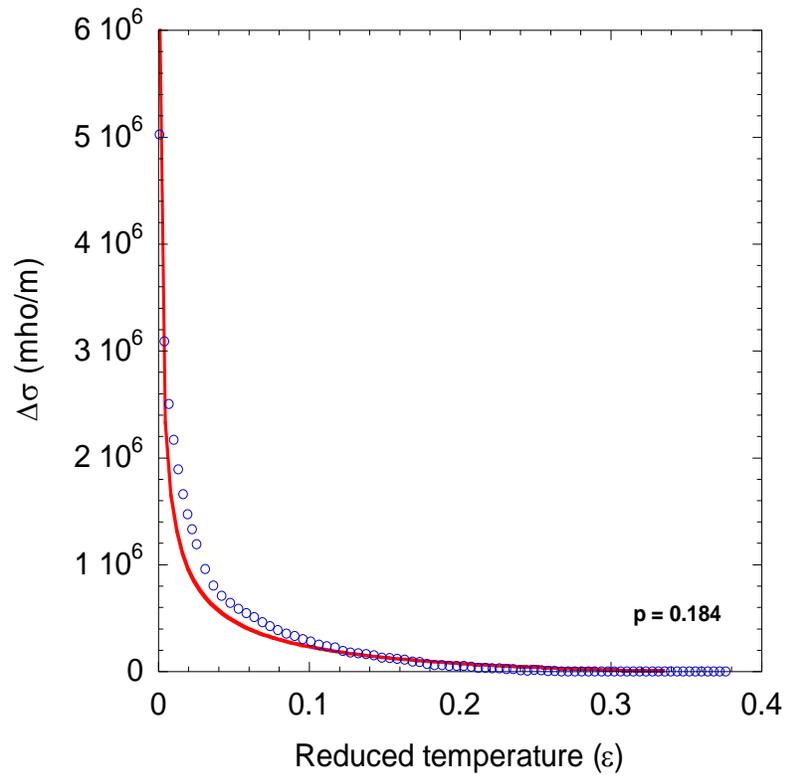